\documentclass[pra,twocolumn,amsmath,amssymb,superscriptaddress,longbibliography]{revtex4-1}
\usepackage{graphicx,amsmath,relsize,epstopdf,upgreek,color,mathtools,bm,mathptmx}

\newcommand{\ket}[1]{\left|{#1}\right>}
\newcommand{\bra}[1]{\left<{#1}\right|}
\newcommand{\rvec}[1]{\pmb{#1}}
\newcommand{\dyadic}[1]{\pmb{#1}}

\newcommand{\D}{\mathrm{d}}
\newcommand{\I}{\mathrm{i}}
\newcommand{\TP}[1]{{#1}^\mathrm{\,\textsc{t}}}
\newcommand{\E}[1]{\mathrm{e}^{\mbox{\footnotesize$#1$}}}

\newcommand{\DET}[1]{\det\!\left\{#1\right\}}

\newcommand{\LAG}[2]{\mathrm{L}_{\,#1}\!\left(#2\right)}

\newcommand{\MEAN}[2]{\mathbb{E}_{#1}\!\left[#2\right]}

\newcommand{\WIGA}[1]{W\!\left(#1\right)}
\newcommand{\QA}[1]{Q\!\left(#1\right)}

\newcommand{\sCRB}{\mathrm{sCRB}}
\newcommand{\sMSE}{\mathrm{sMSE}}
\newcommand{\HERM}[2]{\mathrm{H}_{\,#1}\!\left(#2\right)}

\newcommand{\erfi}[1]{\mathrm{erfi}\!\left(#1\right)}

\begin{document}

\title{Towards optimal quantum tomography with unbalanced homodyning}

\author{Y.~S.~Teo}
\affiliation{BK21 Frontier Physics Research Division, 
	Seoul National University, 08826 Seoul, South Korea}

\author{H.~Jeong}
\affiliation{Center for Macroscopic Quantum Control,
	Seoul National University, 08826 Seoul, South Korea}

\author{L.~L.~S\'{a}nchez-Soto}
\affiliation{Departamento de \'Optica, Facultad de F\'{\i}sica,
	Universidad Complutense, 28040 Madrid, Spain}
\affiliation{Max-Planck-Institut f\"ur  die Physik des Lichts,
	Staudtstra\ss e 2, 91058 Erlangen, Germany}

\begin{abstract}
  Balanced homodyning, heterodyning and unbalanced homodyning are the three well-known sampling techniques used in quantum optics to characterize all possible photonic sources in continuous-variable quantum information theory. We show that for all quantum states and all observable-parameter tomography schemes, which includes the reconstructions of arbitrary operator moments and phase-space quasi-distributions, localized sampling with unbalanced homodyning is always tomographically more powerful (gives more accurate estimators) than delocalized sampling with heterodyning. The latter is recently known to often give more accurate parameter reconstructions than conventional marginalized sampling with balanced homodyning. This result also holds for realistic photodetectors with subunit efficiency. With examples from first- through fourth-moment tomography, we demonstrate that unbalanced homodyning can outperform balanced homodyning when heterodyning fails to do so. This new benchmark takes us one step towards optimal continuous-variable tomography with conventional photodetectors and minimal experimental components.
\end{abstract}


\maketitle

%\tableofcontents

\emph{Introduction}.---In pursuing a secure information age, the successful implementations of state-of-the-art continuous-variable~(CV) quantum information and communication protocols~\cite{Braunstein:2005aa,Ferraro:2005ns,CV2007:aa,Andersen:2010ng,Adesso:2014pm} require precise reconstructions and calibrations of important properties of photonic sources. In the language of phase-space quasi-distributions that completely characterize such sources, these properties---generally the expectation values of quantum observables---can be reconstructed with either physical probabilities of a positive distribution or those derived from some aspects of a (non-singular) quasi-distribution.

There are three sampling methods considered in quantum optics that identify these two main scenarios. The first and arguably the most popular method is \emph{marginalized phase-space sampling} by balanced homodyne detection~(BHOM)~\cite{Yuen:1983ba,Abbas:1983ak, Schumaker:1984qm,Vogel:1989zr,Banaszek:1997ot}, which samples the marginal distributions of the Wigner function defined by quadrature directions. This requires a balanced (1:1) beam splitter, local oscillator (LO), and two photodetectors to measure photocurrent differences at the output. The second method is \emph{delocalized phase-space sampling} executed with heterodyne detection~(HET) that jointly measures complementary quadrature operators~\cite{Arthurs:1965al,Yuen:1982hh,Arthurs:1988aa,Martens:1990al,Martens:1991aa,Stenholm:1992ps,Raymer:1994aj,Trifonov:2001up,Werner:2004as}. This technique randomly samples the whole phase space according to the Husimi function, and usually involves a more sophisticated setup of three balanced beam splitters, LO and four photodetectors to realize such a double-BHOM scheme. The third sampling method of focus here is \emph{localized phase-space sampling} with unbalanced homodyne detection~(UHOM)~\cite{Wallentowitz:1996qo,Banaszek:1996ng,Opatrny:1997qo,Opatrny:1997mu,Wallentowitz:2012aa,Kuhn:2016cm,Paris:1996do}, which measures displaced Fock states using a highly-transmissive beam splitter, LO and two photodetectors such that the Wigner function can be directly reconstructed through the parity-operator measurement. With common photodetectors that have no photon-number resolution capabilities, this method samples the Husimi function by counting ``no-click'' events at the transmission arm of the signal. The displacement operation by the unbalanced beam splitter then guarantees coherent-state measurements of specified amplitudes, which data follow a binomial distribution characterized by the Husimi function at each amplitude.

The understanding of the parameter reconstruction accuracies for all sampling methods holds a fundamental link to the tomographic power of quantum measurement schemes. There exist a plethora of articles~\cite{Yuen:1982hh,Trifonov:2001up,Werner:2004as,Kiesel:2012sm} that investigated variances and measurement uncertainties, which supply information about important statistical behaviors of parameter estimators. For the purpose of analyzing tomographic power, optimality analysis for true-parameter reconstructions is in order. Recently, the relationship between the Haar-averaged Cram{\'e}r--Rao bound for state estimation and the permutation group was studied in \cite{Rehacek:2015qm,Dominik:2016up}. In~\cite{Rehacek:2015qp,Muller:2016da,Teo:2017aa}, we systematically analyzed the tomographic power of both BHOM and HET using the notions of the Fisher information and Cram{\'e}r--Rao bound for moment estimation and found that the latter gives higher reconstruction accuracies for typically interesting states, with Gaussian states being one important class in CV quantum information processing~\cite{Lorenz:2004aa, Ferraro:2005ns,Lance:2005aa,Rehacek:2009ys,Scarani:2009cq,Weedbrook:2012ag}. This provided irrefutable evidence of tomographic differences in parameter reconstruction for the Wigner and Husimi representations, despite their equivalence in state representation.

It will be shown here that for every used reconstruction datum, localized sampling with UHOM is always tomographically more powerful than delocalized sampling with HET for \emph{any} type of observable-parameter tomography. This benefit originates from the statistical nature of the UHOM data collected at each phase-space value. We shall demonstrate that this effect can even result in a superior tomographic power over BHOM in some cases where HET is inferior. These two main results are analyzed for first- through fourth-order moment tomography with Gaussian and Fock states.

\emph{Parameters and tomographic power}.---For a more concrete concept of comparing different measurement schemes, we consider the statistical mean squared error (MSE) $\MEAN{}{(\widehat{\rvec{q}}-\rvec{q})^2}$ for any column $\rvec{q}$ of parameters and its estimator $\widehat{\rvec{q}}$. This accuracy measure is a function of both the measurement and data for $\widehat{\rvec{q}}$. We consider \emph{observable parameters} of the kind $\rvec{q}=\left<\rvec{V}\right>$ for an arbitrary column $\rvec{V}$ of observables describing some list of quantum properties (which is always a function of the position $X$ and momentum $P$ operators~\cite{Englert:QM2}), where equivalently~\cite{Cahill:1969qd}
\begin{equation}
\rvec{q}=\int\frac{(\D\alpha')}{\pi}\WIGA{\alpha'}\rvec{v}_\textsc{w}(\alpha')=\int\frac{(\D\alpha')}{\pi}\QA{\alpha'}\rvec{v}_\textsc{p}(\alpha')
\label{eq:q}
\end{equation}
is the phase-space average of either the Glauber-Sudarshan function $\rvec{v}_\textsc{p}(\alpha)$ or the Wigner function $\rvec{v}_\textsc{w}(\alpha)=2\int(\D\alpha')\,\E{-2|\alpha-\alpha'|^2}\rvec{v}_\textsc{p}(\alpha')/\pi$ for $\rvec{V}$, respectively with the Husimi and Wigner functions for the state $\rho$ [$0\leq\QA{\alpha}\leq1$, $-2\leq\WIGA{\alpha}\leq2$ according to the definitions for Eq.~\eqref{eq:q}]. Each point $(x,p)$ in phase space is expressed by $\alpha=(x+\I p)/2$, and $(\D\alpha)=\D x\,\D p/2$. Equation~\eqref{eq:q} covers all the interesting tomography problems. For instance, in second-moment tomography where $\rvec{V}$ consists of symmetrically-ordered products of $X$ and $P$, then $\rvec{v}_\textsc{w}(\alpha')=\TP{(x'^2,x'p',p'^2)}$ and $\rvec{v}_\textsc{p}(\alpha')=\TP{(x'^2-1/2,x'p',p'^2-1/2)}$. As another example, if one is interested in Wigner function reconstruction, then $\rvec{v}_\textsc{w}(\alpha')=\delta(\rvec{\alpha}-\alpha')$ and $\rvec{v}_\textsc{p}(\alpha')$ is the kernel for the Gaussian deconvolution. For an $s$-ordered quasi-distribution, $\rvec{v}_\textsc{w}(\alpha')$ and $\rvec{v}_\textsc{p}(\alpha')$ are the relevant $s$-ordered kernels.

\begin{figure}[t]
	\centering
	\includegraphics[width=1\columnwidth]{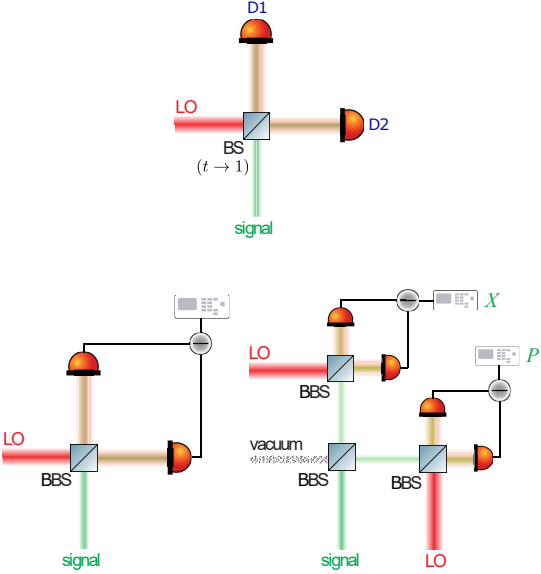}
	\caption{\label{fig:Fig1}Schema for the (a)~UHOM, (b)~BHOM and (c)~HET setups. The UHOM scheme consists only of one beam splitter (BS) that is almost perfectly transmissive (transmission amplitude $t\rightarrow1$) and two photodetectors D1 and D2, where only data corresponding to vacuum-state measurement with D1 are used to estimate the Husimi function, which is a significant experimental simplification compared to HET, which requires three balanced BSs (BBS) and four photodetectors to randomly sample the Husimi function.}
\end{figure}

The detection schemes for all three sampling methods have very different kinds of data. The data sample size $N$ for BHOM is the total number of marginalized Wigner data points defined by the sampled LO phases and real voltages. For HET and UHOM, $N$ is the total number of randomly sampled phase-space values, but since only the ``no-click'' data for the D1 photodetector (out of a fixed number of sampling events per $\alpha$) at the transmitted arm for the signal~(see Fig.~\ref{fig:Fig1}) are used in the reconstruction, $N$ becomes a sum of binomial random integers, and is itself random. To make a fair comparison of the three schemes, the well-known (scaled) Cram{\'e}r--Rao bound $\sCRB=\min_{\widehat{\rvec{q}}}\{\MEAN{}{N(\widehat{\rvec{q}}-\rvec{q})^2}\}$ is a good measure for the tomographic power of the measurement. This scaled measure consistently weights each experiment with its total sample size to average away the data aspect, and is minimized over all conceivable reconstruction strategies for $\widehat{\rvec{q}}$ of some given data type. A smaller sCRB implies a greater tomographic power. For BHOM and HET, $N$ is usually a fixed constant, so that the sCRB turns into the familiar MSE per reconstruction datum. For sufficiently large coherent-state data $N$ and densely sampled phase-space points, one can show that the MSE for UHOM goes as the average shot-noise limit \mbox{$\left(\propto 1/\MEAN{}{N}\right)$}, which again reminds us that the accuracy of $\widehat{\rvec{q}}$ varies only with the \emph{used} reconstruction data sample size as always. It then follows that $\min_{\widehat{\rvec{q}}}\{\MEAN{}{N(\widehat{\rvec{q}}-\rvec{q})^2}\}=\MEAN{}{N}\min_{\widehat{\rvec{q}}}\{\MEAN{}{(\widehat{\rvec{q}}-\rvec{q})^2}\}$. This means that while the comparisons of the sampling methods are made by scaling away the used reconstruction data, it should not matter whether this scaling is done for every experiment or with an overall average data cost for all the experiments. Any physically meaningful definition of the tomographic power should be invariant under such a technical variation.

\emph{Main results}.---Both the delocalized (HET) and localized (UHOM) phase-space sampling methods share a common trait: their data $N=\sum_ln_l$ directly reconstruct the Husimi function $\QA{\alpha}$: $n_l/\sum_ln_l\approx(\updelta\alpha)\QA{\alpha_l}/\pi$ at a sampled $\alpha=\alpha_l$ for some small pre-chosen area $(\updelta\alpha)$ of the sampled discretized phase-space. Therefore the Husimi representation of $\rvec{q}$ in \eqref{eq:q} invites an estimator of the form $\widehat{\rvec{q}}=\sum_ln_l\,\rvec{v}_\textsc{p}(\alpha_l)/\sum_ln_l$. Furthermore, it can be shown that such a sample average estimator, for these HET and UHOM data, follows a multivariate Gaussian distribution in the limit of large $N$ with the correct mean $\rvec{q}$ and data covariance, so that $\widehat{\rvec{q}}$ achieves the sCRB asymptotically. After a proper statistical analysis for $N\gg1$ and a densely sampled $\alpha_l$s, we have the intuitively simple expressions (see App.~\ref{app:deriv})
\begin{align}
\sCRB_\textsc{het}=&\,\int\dfrac{(\D\alpha')}{\pi}\QA{\alpha'}\left[\rvec{v}_\textsc{p}(\alpha')-\rvec{q}\right]^2\,,\nonumber\\
\sCRB_\textsc{uhom}=&\,\int\dfrac{(\D\alpha')}{\pi}\QA{\alpha'}\left[1-\QA{\alpha'}\right]\left[\rvec{v}_\textsc{p}(\alpha')-\rvec{q}\right]^2\,.
\label{eq:scrb1}
\end{align}
By inspection, since $\QA{\alpha}\left[1-\QA{\alpha}\right]\leq\QA{\alpha}$ for any $\rho$, we immediately find that $\sCRB_\textsc{uhom}<\sCRB_\textsc{het}$.
This first general result has important physical implications. It shows that localized sampling always reduces the magnitude of the phase-space distribution \emph{via} the binomial deformation $\QA{\alpha}\rightarrow\QA{\alpha}\left[1-\QA{\alpha}\right]$. This leads to a smaller combined reconstruction variance per datum for any list of parameters $\rvec{q}$ relative to HET. In practice, the tomographic advantages of localized binomial phase-space sampling is realized with only a replacement of the balanced BS with a highly-transmissive BS, which is a minor adjustment of the BHOM setup in Fig.~\ref{fig:Fig1}(b). For realistic photodetectors of efficiency $0<\eta\leq1$, using the definition $p(\alpha,\eta)=\left<\bm{:}\E{-\eta(a^\dagger-\alpha)(a-\alpha)}\bm{:}\right>\leq1$ where $a$ is the usual photonic ladder operator and $\bm{:}\,\cdot\,\bm{:}$ denotes operator normal ordering, the sampled probabilities with HET are given by $\eta p(\alpha,\eta)$ while the binomial probability for ``no-click'' events at photodetector D1 for UHOM is $p(\alpha,\eta)$~\cite{Wallentowitz:1996qo}. These give the more realistic bounds (see App.~\ref{app:real})
\begin{align}
\sCRB'_\textsc{het}=&\,\eta\int\dfrac{(\D\alpha')}{\pi}p(\alpha',\eta)\left[\rvec{v}_\textsc{p}(\alpha')-\rvec{q}\right]^2\,,\nonumber\\
\sCRB'_\textsc{uhom}=&\,\eta\int\dfrac{(\D\alpha')}{\pi}p(\alpha',\eta)\left[1-p(\alpha',\eta)\right]\left[\rvec{v}_\textsc{p}(\alpha')-\rvec{q}\right]^2
\label{eq:scrb2}
\end{align}
that satisfy $\sCRB'_\textsc{uhom}<\sCRB'_\textsc{het}$~\footnote{The bounds should diverge as $\eta\rightarrow0$, which means that one cannot swap this limit with the phase-space integration}.

\begin{figure}[t]
	\centering
	\includegraphics[width=1\columnwidth]{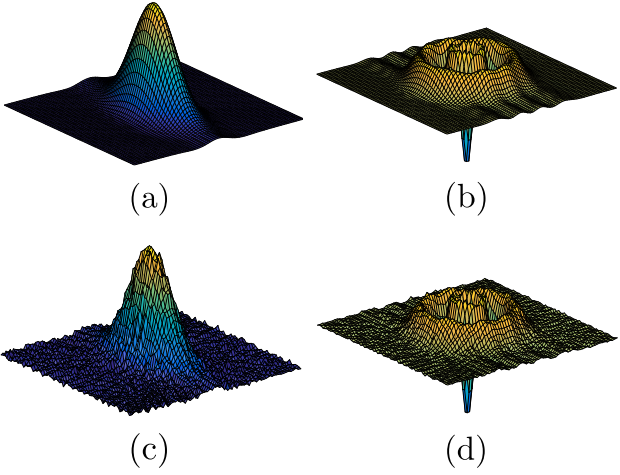}
	\caption{\label{fig:Fig2}Wigner functions of (a,c)~a squeezed Gaussian state and (b,d)~a Fock state of $n=3$ reconstructed with a truncated invR for BHOM based on (a,b)~perfect data and (c,d)~noisy data. The wriggles of the reconstructed functions that come from truncations to a 50-dimensional Hilbert subspace, even for the case of perfect data, can lead to significant deviations from the true $\rvec{q}$.}
\end{figure}

\begin{figure}[t]
	\centering
	\includegraphics[width=1\columnwidth]{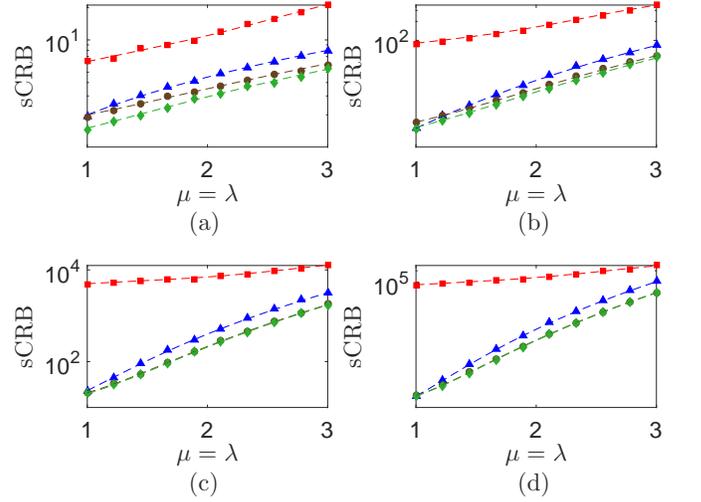}
	\caption{\label{fig:Fig3}Plots for the (a)~first-, (b)~second-, (c)~third- and (d)~fourth-moment reconstruction of a Gaussian state of $1\leq\mu=\lambda\leq3$, in which the $\sCRB$ of BHOM (square markers and curve), BHOMOPT (triangular markers and curve), HET (circular markers and curve), and UHOM (diamond markers and curve) are illustrated. The instability and sensitivity of the invR with the BHOM marginalized sampling strategy is clear in the plots, which behavior also depends on the truncated Hilbert space. Evidently, UHOM exhibits a more superior tomographic power than HET, BHOM and BHOMOPT. The dashed curves represent theory derived from \eqref{eq:scrb1} and \eqref{eq:scrb2}, whereas the markers are computed with Monte Carlo simulated data of the CV experiments for a $50$-dimensional Hilbert space. For the purpose of illustrating the results, we take $\eta=1$ for simplicity.}
\end{figure}

For an arbitrary $\rvec{V}$ that is a complicated function of $X$ and $P$, the general recipe for $\widehat{\rvec{q}}$ with BHOM data is to adopt the Wigner representation in \eqref{eq:q} and estimate $\WIGA{\alpha}$ by an application of the inverse Radon transform (invR) to the BHOM probabilities. Upon denoting the invR kernel by $\mathcal{R}^{-1}_\alpha(x_\vartheta,\vartheta)=\int\D k|k|\exp(\I k(x\cos\vartheta+p\sin\vartheta-x_\vartheta))$ for a given LO phase $\vartheta$ and voltage $x_\vartheta$, the corresponding estimator for the BHOM data is given by $\widehat{\rvec{q}}=\sum_{l,j,k}\mathcal{R}^{-1}_{\alpha_l}(x_j,\vartheta_k)n_{jk}\,\rvec{v}_\textsc{w}(\alpha_l)/\sum_{l,j,k}\mathcal{R}^{-1}_{\alpha_l}(x_j,\vartheta_k)n_{jk}$, where $n_{jk}/\sum_jn_{jk}$ estimates the BHOM probability $\D x\, p(x_j,\vartheta_k)$. The tomographic power of BHOM for this general recipe with invR (only one kind of estimator considered here) is measured by
\begin{align}
\sCRB_\textsc{bhom}=&\,\int\dfrac{(\D\alpha')}{\pi}\int\dfrac{(\D\alpha'')}{\pi}w_{\alpha',\alpha''}\nonumber\\
&\qquad\qquad\qquad\quad\times[\rvec{v}_\textsc{w}(\alpha')-\rvec{q}]\bm{\cdot}[\rvec{v}_\textsc{w}(\alpha'')-\rvec{q}]\,,\nonumber\\
w_{\alpha',\alpha''}=&\,\int_{(\pi)}\frac{\D\vartheta}{2\pi}\int\D x'_\vartheta\int\D x''_\vartheta\, \mathcal{R}^{-1}_{\alpha'}(x'_\vartheta,\vartheta)\,\mathcal{R}^{-1}_{\alpha''}(x''_\vartheta,\vartheta)\nonumber\\
&\,\times\left[p(x'_\vartheta,\vartheta)\delta(x'_\vartheta-x''_\vartheta)-p(x'_\vartheta,\vartheta)p(x''_\vartheta,\vartheta)\right]\,.
\label{eq:scrb2}
\end{align}
In practice, the estimation of the Wigner function $\WIGA{\alpha}$ is done in a truncated Hilbert space. As such, a direct application of the invR on the measured BHOM probabilities typically gives rise to $\WIGA{\alpha}$ with truncation phase-space wriggles that are otherwise absent in the infinite-dimensional limit. Together with the high sensitivity of invR to statistical fluctuations, $\sCRB_\textsc{bhom}$ is in general greater than either $\sCRB_\textsc{uhom}$ or $\sCRB_\textsc{het}$ (see Fig.~\ref{fig:Fig2}). Thus, for general parameters where $\rvec{V}$ is a complicated function of $X$ and $P$, UHOM is the best option. Although it is known that the maximum-likelihood method can reduce such reconstruction instabilities~~\cite{Banaszek:1999ml,Banaszek:1999ap}, analytical tomographic studies of such a nonlinear numerical method still form an open problem.

Certainly, a much more expedient and trusted way to estimate $\rvec{q}$ (referred to as the BHOMOPT strategy) when $\rvec{V}$ is a simple function of $X$ and $P$ is a direct and optimized data-processing strategy of the measured voltage values for every LO phase $\vartheta$ such that $\rvec{q}$ can be efficiently reconstructed without having to go through any formal invR. For instance, in moment tomography \cite{Rehacek:2009ys,Muller:2016da,Teo:2017aa}, the entries of $\rvec{q}$ are linearly related to the moments of the quadrature operator, $\left<X^m_\vartheta\right>=\left<(X\cos\vartheta+P\sin\vartheta)^m\right>$, sampled by BHOM. Therefore, $\sCRB_\textsc{bhomopt}$ for any state using this improved reconstruction strategy can be obtained through the Fisher information of the homodyne parameter $\left<X^m_\vartheta\right>$. The theory for this was developed in~\cite{Teo:2017aa}. It shall be shown that in practice, UHOM is tomographically more powerful than all other methods for moment tomography of interesting states, which forms the second main result.

\begin{figure}[t]
	\centering
	\includegraphics[width=1\columnwidth]{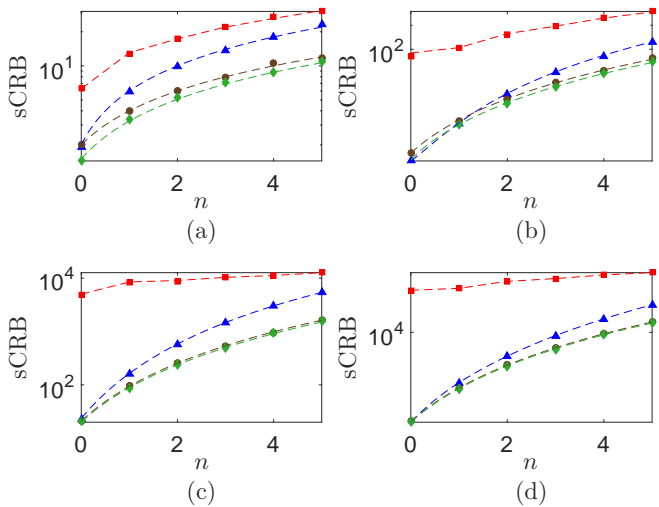}
	\caption{\label{fig:Fig4}Plots of $\sCRB$ for the Fock states of $0\leq n\leq5$. All specifications follow those of Fig.~\ref{fig:Fig3}. For each BHOM plot, the dashed curve joins the six theoretically calculated numerical values that match the square markers.}
\end{figure}

\emph{Moment-tomography analysis}.---We demonstrate the tomographic power of UHOM with moment tomography of orders $m=1$ through $m=4$. In particular, we study symmetrically-ordered operator moments of $X$ and $P$ that appear naturally in high-order operator covariances. As examples, we consider two classes of quantum states. The first example is the class of (centralized) Gaussian states described by a covariance matrix of eigenvalues $\mu\lambda/2$ and $\mu/(2\lambda)$, where $\mu$ is related to the thermal mean photon number or temperature and $\lambda$ describes the squeezing strength. For simplicity, we set $\mu=\lambda$, which approximately models strongly-squeezed sources with accompanying excess noise associated to the anti-squeezed quadrature due to realistic experimental imperfections~(refer for example to Ref.~\cite{Muller:2012ys}). The second example is the class of Fock states of $n$ photon numbers which are arguably the most non-Glauber-Sudarshan-representable states.  Even for these states, there is in general no known explicit expressions for $\sCRB_\textsc{bhom}$ in the state parameters and numerical techniques are needed to calculate its values. The expressions for the sCRBs are listed in App.~\ref{app:momtom}.

Figures~\ref{fig:Fig3} and \ref{fig:Fig4} present the findings for these states. As intuitively expected, the more direct BHOMOPT reconstruction of the moments is always (exponentially) better than estimating the Wigner function with BHOM. Even then, this improved strategy still often underperforms in comparison to HET and UHOM. For the Gaussian states, when $m=1$ or 3, marginalized sampling with BHOM and BHOMOPT give the worst tomographic performance. Localized sampling with the UHOM strategy generates the most accurate estimators per reconstruction datum, and delocalized sampling with HET is second best. When $m=2$ or 4, BHOMOPT beats HET respectively for $\mu\lesssim1.262$ and $\mu\lesssim1.017$, after which HET catches up in tomographic power, whereas UHOM ranks the top in the respective ranges $\mu\gtrsim1.04$ and $\mu\gtrsim1.004$. Likewise for the Fock states, both HET and  UHOM, beat BHOM and BHOMOPT for all $n$ values and $m=1,3$. When $m=2$ or 4, BHOMOPT initially outperforms HET for the vacuum state (and also the $n=1$ state for $m=2$) and subsequently becomes inferior to HET. UHOM on the other hand is superior to all methods in tomographic power for all $n>0$. 

That HET surpasses BHOMOPT for the $m=1$ case for \emph{any} state is a consequence of the Heisenberg-Robertson-Schr{\"o}dinger uncertainty relation~\cite{Teo:2017aa}. The limiting case where the two methods give identical sCRBs is when the state is of minimum uncertainty. Yet, UHOM is able to overcome this limit owing to the binomial variances. For the vacuum ($\mu=1$ or $n=0$), both UHOM and BHOMOPT are almost identical in power ($\sCRB_\textsc{uhom}/\sCRB_\textsc{bhomopt}=33/32\approx1.031$ and $9879/9856\approx1.002$ for $m=2$ and $4$) within experimental error margins. This justifies the use of UHOM essentially for all these states.

\emph{Conclusion}.---We have first proven that, for every \emph{used} reconstruction datum, localized phase-space sampling with unbalanced homodyning \emph{always} beats delocalized phase-space sampling with heterodyning in tomographic power measured by the scaled Cram{\'e}r--Rao for any quantum state and general multivariate observable-parameter tomography. The reason is attributed to the binomial nature of unbalanced homodyne data, which enhances the resolution of Husimi-function reconstruction with fewer experimental components. We next demonstrated that for the Gaussian states and Fock states, localized sampling almost always beats marginalized sampling with balanced homodyning in moment tomography, except for the vacuum where both methods are practically equals. These findings shed light on the general performances of sampling methods in continuous-variable tomography.

\begin{acknowledgements}
The author thanks D.~Ahn, Z.~Hradil, and J.~{\v R}eh{\'a}{\v c}ek for illuminating discussions. This work is financially supported by the BK21 Plus Program (21A20131111123) funded by the Ministry of Education (MOE, Korea) and National Research Foundation of Korea (NRF), the NRF grant funded by the Korea government (MSIP) (Grant No. 2010-0018295), the KIST Institutional Program (Project No. 2E26680-16-P025), the European Research Council (Advanced Grant PACART), as well as the Spanish MINECO (Grant FIS2015-67963-P).
\end{acknowledgements}

\appendix

\section{Derivations of Eqs.~(2) and (4)}
\label{app:deriv}

To arrive at the expression for $\sCRB_\textsc{het}$, we first note that since $N$ is fixed for HET, it is sufficient to use the standard formula $\MEAN{}{n_ln_{l'}}=Np_l\delta_{l,l'}+N(N-1)p_lp_{l'}$ for the binned multinomial HET data with $p_l\approx(\D\alpha)\QA{\alpha_l}/\pi$. For UHOM, we would need the averages
\begin{align}
\MEAN{}{\dfrac{n_l}{N}}=&\,\dfrac{p_l}{\sum_lp_l}-\dfrac{\sigma_l^2}{(\sum_lp_l)^2}+\dfrac{p_l}{(\sum_lp_l)^3}\sum_{l'}\sigma_{l'}^2\,,\label{eq:frac_uhom}\\
\MEAN{}{\dfrac{n_ln_{l'}}{N^2}}=&\,\dfrac{\sigma_l^2\delta_{l,l'}+p_lp_{l'}}{(\sum_lp_l)^2}-\dfrac{2\left(p_l\sigma^2_{l'}+p_{l'}\sigma^2_l\right)}{(\sum_lp_l)^3}\label{eq:frac_uhom2}\\
&\,+\dfrac{3p_lp_{l'}}{(\sum_lp_l)^4}\sum_{l''}\sigma^2_{l''}\,,
\end{align}
where here $p_l$ is instead equal to $\QA{\alpha_l}$, $\sigma_l^2=p_l(1-p_l)/N_0$, and $N_0$ is the total number of detection events for each $\alpha_l$.

The averages of data-ratios in Eqs.~\eqref{eq:frac_uhom} and \eqref{eq:frac_uhom2} can be straightforwardly derived by starting with this easy but crucial integral identity
\begin{equation}
\dfrac{1}{A}=-\I\int^\infty_0\D t\,\E{\I t A-\epsilon t}\bigg|_{\epsilon=0}
\label{eq:recip_int}
\end{equation}
for any non-zero $A$. We can then rewrite the UHOM ratio averages as
\begin{align}
\MEAN{}{\dfrac{f_l}{\sum_lf_l}}=&\,-\int^\infty_0\D t\dfrac{\partial}{\partial\lambda_l}\MEAN{}{\E{\I \sum_l\lambda_lf_l}}\bigg|_{\lambda_l=t}\nonumber\\
\MEAN{}{\dfrac{f_lf_{l'}}{(\sum_lf_l)^2}}=&\,\int^\infty_0\D t\int^\infty_0\D t'\,\dfrac{\partial}{\partial\lambda_l}\dfrac{\partial}{\partial\lambda_{l'}}\MEAN{}{\E{\I\sum_l\lambda_lf_l}}\bigg|_{\lambda_l=t+t'}
\end{align}
after an equivalent renormalization $f_l=n_l/N_0$ for notational simplicity. The central object to be evaluated is thus the characteristic function $\MEAN{}{\E{\I\sum_l\lambda_lf_l}}$. As the data for distinct $\alpha_l$ are statistically independent, we may proceed with the decomposition
\begin{equation}
\MEAN{}{\E{\I\sum_l\lambda_lf_l}}=\prod_{l}\MEAN{}{\E{\I\lambda_lf_l}}\,,
\end{equation}
where central limit theorem gives
\begin{equation}
\MEAN{}{\E{\I\lambda_lf_l}}=\E{-\frac{1}{2}\lambda_{l}^2\sigma_l^2+\I\lambda_lp_l}
\end{equation}
for $N_0\gg1$, and the corresponding integrals
\begin{equation}
\MEAN{}{\dfrac{f_l}{\sum_lf_l}}=-\int^\infty_0\D t\,\E{-at^2+\I bt}\,(\I p_l-t \sigma_l^2)\,,
\end{equation}
and
\begin{align}
\MEAN{}{\dfrac{f_l\,f_{l'}}{\left(\sum_lf_l\right)^2}}=&\,\int^\infty_0\D t\,\int^\infty_0\D t'\,\E{-a(t+t')^2+\I b(t+t')}\nonumber\\
&\qquad\qquad\quad\times\left[c_1(t+t')^2-\I c_2(t+t')-c_3\right]\,,
\end{align}
where the parameters are now $\sigma_l^2=p_l(1-p_l)$, $a=\sum_l\sigma^2_l/2$, $b=\sum_lp_l$, $c_1=\sigma^2_l\sigma^2_{l'}$, $c_2=p_l\sigma^2_{l'}+p_{l'}\sigma^2_l$ and $c_3=p_lp_{l'}+\delta_{l,l'}\sigma^2_l$. The answers to these integrals involve the imaginary error function $\erfi{b/2\sqrt{a}}$, and in the limit of large sample size where $b\gg 2\sqrt{a}$, it turns out that the asymptotic expansion
\begin{equation}
\sqrt{\pi}\,\E{-\frac{b^2}{4a}}\,\erfi{\frac{b}{2\sqrt{a}}}\approx\frac{24 a^{5/2}}{b^5}+\frac{4 a^{3/2}}{b^3}+\frac{2 \sqrt{a}}{b}
\label{eq:erfi}
\end{equation}
gives Eqs.~\eqref{eq:frac_uhom} and \eqref{eq:frac_uhom2} for sufficiently dense sampling $(0<b\gg1)$. The integral expressions for $\sCRB_\textsc{het}$ and $\sCRB_\textsc{uhom}$ are the acquired limits of such a dense sampling.

The important technical statement $\min_{\widehat{\rvec{q}}}\{\MEAN{}{N(\widehat{\rvec{q}}-\rvec{q})^2}\}=\MEAN{}{N}\min_{\widehat{\rvec{q}}}\{\MEAN{}{(\widehat{\rvec{q}}-\rvec{q})^2}\}$ is then easily proven with the additional statistical identity
\begin{equation}
\MEAN{}{\dfrac{n_ln_{l'}}{N}}=\dfrac{\sigma_l^2\delta_{l,l'}+p_lp_{l'}}{\sum_lp_l}-\dfrac{p_l\sigma^2_{l'}+p_{l'}\sigma^2_l}{(\sum_lp_l)^2}+\dfrac{p_lp_{l'}}{(\sum_lp_l)^3}\sum_{l''}\sigma^2_{l''}\,,
\label{eq:extra}
\end{equation}
which can also be derived with 
\begin{align}
\MEAN{}{\dfrac{f_l\,f_{l'}}{\sum_lf_l}}=&\,\int^\infty_0\D t\,\E{-at^2+\I bt}\left[c_1t^2-\I c_2t-c_3\right]
\end{align}
of the same parameters defined above after a similar calculation.

To get $\sCRB_\textsc{bhom}$, we need the data-ratio averages
\begin{align}
\MEAN{}{\dfrac{n_{jk}}{\mathcal{N}}}=&\,\dfrac{p_{jk}}{b}+\dfrac{2ap_{jk}}{b^3}-\dfrac{w_{jk}}{b^2}\,,\label{eq:frac_bhom}\\
\MEAN{}{\dfrac{n_{jk}\,n_{j'k'}}{\mathcal{N}^2}}=&\,\dfrac{p_{jk}p_{j'k'}+\delta_{j,j'}\Sigma_{jkk'}}{b^2}+\dfrac{6ap_{jk}p_{j'k'}}{b^4}\nonumber\\
&\,-\dfrac{2}{b^3}(p_{jk}w_{j'k'}+p_{j'k'}w_{jk})\label{eq:frac_bhom2}
\end{align}
that hold when BHOM sampling is sufficiently dense ($|b|\gg1\implies N\gg1$), with $\mathcal{N}=\sum_l\sum^{n_\vartheta}_{j=1}\sum^{n_x}_{k=1}\mathcal{R}^{-1}_{ljk}n_{jk}$,  $p_{jk}=\D x_k\,p(x_k,\vartheta_j)$, $\widetilde{N}=\sum^{n_x}_{k=1}n_{jk}$, $\Sigma_{jkk'}=(p_{jk}\delta_{k,k'}-p_{jk}p_{jk'})/\widetilde{N}$, $w_{jk}=\sum_{k',l'}\mathcal{R}^{-1}_{l'jk'}\Sigma_{jkk'}$, $a=\sum_{j,k,l}\mathcal{R}^{-1}_{ljk}w_{jk}/(4n_\vartheta)$ and $b=\sum_{j,k,l}\mathcal{R}^{-1}_{ljk}p_{jk}/(2n_\vartheta)$. The averages in \eqref{eq:frac_bhom} and \eqref{eq:frac_bhom2} can be verified with \eqref{eq:recip_int}, which yields 
\begin{align}
\MEAN{}{\dfrac{f_{jk}}{\sum_{l,j,k}\mathcal{R}^{-1}_{ljk}f_{jk}}}=&\,-\I\int^\infty_0\D t\,\,\MEAN{}{f_{jk}\E{\I t\sum_{l,j,k}\mathcal{R}^{-1}_{ljk}f_{jk}}}\\
=&\,-\int^\infty_0\D t\dfrac{\partial}{\partial\lambda_{ljk}}\MEAN{}{\E{\I \sum_{l,j,k}\lambda_{ljk}f_{jk}}}\bigg|_{\lambda_{ljk}=t\mathcal{R}^{-1}_{ljk}}
\end{align}
and
\begin{align}
&\,\MEAN{}{\dfrac{f_{jk}\,f_{j'k'}}{\left(\sum_{l,j,k}\mathcal{R}^{-1}_{ljk}f_{jk}\right)^2}}\nonumber\\
=&\,-\int^\infty_0\D t\int^\infty_0\D t'\,\,\MEAN{}{f_{jk}f_{j'k'}\E{\I (t+t')\sum_{l,j,k}\mathcal{R}^{-1}_{ljk}f_{jk}}}\nonumber\\
=&\,\int^\infty_0\D t\int^\infty_0\D t'\,\dfrac{\partial}{\partial\lambda_{ljk}}\dfrac{\partial}{\partial\lambda_{l'j'k'}}\MEAN{}{\E{\I\sum_{l,j,k}\lambda_{ljk}f_{jk}}}\bigg|_{\lambda_{ljk}=(t+t')\mathcal{R}^{-1}_{ljk}}\,,
\end{align}
after a renormalization $f_{jk}=n_{jk}/\widetilde{N}$ with the constant $\widetilde{N}$.

We again realize that the data for distinct $\vartheta_j$ are statistically independent, which means that 
\begin{equation}
\MEAN{}{\E{\I\sum_{l,j,k}\lambda_{ljk}f_{jk}}}=\prod^{n_\vartheta}_{j=1}\MEAN{}{\E{\I\sum_{l,k}\lambda_{ljk}f_{jk}}}
\end{equation}
decomposes into the independent characteristic functions. Because the binned data $\{n_{jk}\}_k$ for every $j$ follows a multinomial distribution defined by the BHOM quantum probabilities $\sum_kp_{jk}=1$, in the limit of large $\widetilde{N}$, the column $\rvec{f}_j=(f_{jk})$ follows a Gaussian distribution of mean $\rvec{p}_j$ and covariance matrix $\left[\mathrm{diag}(\rvec{p}_j)-\rvec{p}_j\rvec{p}_j\right]/\widetilde{N}$, so that
\begin{equation}
\MEAN{}{\E{\I\sum_{l,k}\lambda_{ljk}f_{jk}}}=\E{-\frac{1}{2}\sum_{l,l',k,k'}\lambda_{ljk}\lambda_{l'jk'}\Sigma_{jkk'}+\I\sum_{l,k}\lambda_{ljk}p_{jk}}
\end{equation}
according to the central limit theorem. After the differentiations, we have
\begin{equation}
\MEAN{}{\dfrac{f_{jk}}{\sum_{l,j,k}\mathcal{R}^{-1}_{ljk}f_{jk}}}=-\int^\infty_0\D t\,\E{-at^2+\I bt}\,(\I p_{jk}-t w_{jk})\,,
\end{equation}
and
\begin{align}
\MEAN{}{\dfrac{f_{jk}\,f_{j'k'}}{\left(\sum_{l,j,k}\mathcal{R}^{-1}_{ljk}f_{jk}\right)^2}}=&\,\int^\infty_0\D t\,\int^\infty_0\D t'\,\E{-a(t+t')^2+\I b(t+t')}\nonumber\\
&\quad\times\left[c_1(t+t')^2-\I c_2(t+t')-c_3\right]\,,
\end{align}
where $c_1=w_{jk}w_{j'k'}$, $c_2=p_{jk}w_{j'k'}+p_{j'k'}w_{jk}$ and $c_3=p_{jk}p_{j'k'}+\delta_{j,j'}\Sigma_{jkk'}$. From the asymptotic formula in \eqref{eq:erfi} that holds for $N\gg1$, we arrive at the results in Eqs.~\eqref{eq:frac_bhom} and \eqref{eq:frac_bhom} up to $O(1/N)$ as long as phase-space sampling is sufficiently dense or $|b|\gg1$.

It remains to show that the sMSEs are indeed sCRBs. For this, we may invoke the central limit theorem for the respective sums of random variables, the key structures of the HET and UHOM estimators, to prove that $\sMSE\rightarrow\sCRB$ in the limit of dense sampling and/or accurate sampling. The especially important verification that the summands for the UHOM estimator are asymptotically independent is given in the next section. For HET, independence is clear.

\section{Asymptotic independence of UHOM random variables}
\label{app:indrv}

Let $z_l\equiv n_l/N$. Then for any two UHOM random variables $z_l$ and $z_{l'}$, the one-dimensional version of Kac's theorem~\cite{Applebaum:2005qi} states that if we can show that the two-dimensional characteristic function
\begin{equation}
\MEAN{}{\exp(\I(k_lz_l+k_{l'}z_{l'}))}=\MEAN{}{\exp(\I k_lz_l)}\,\MEAN{}{\exp(\I k_{l'}z_{l'})}
\end{equation}
satisfies this decomposition rule for all real $k_l$ and $k_{l'}$, then $z_l$ and $z_{l'}$ are statistically independent, and \emph{vice versa}. This is equivalent to showing that $\MEAN{}{z_l^mz_{l'}^{m'}}=\MEAN{}{z_l^m\vphantom{z_{l'}^{m'}}}\MEAN{}{z_{l'}^{m'}}$ for any $l$, $l'\neq l$, $m$ and $m'$.

For sufficiently large $N_0$, the first-order Taylor expansion in $n_l$ about $N_0p_l$ well approximates
\begin{equation}
\dfrac{1}{(\sum_ln_l)^{m+m'}}\approx\dfrac{(m+m'+1)N_0\sum_lp_l-(m+m')\sum_ln_l}{(N_0\sum_lp_l)^{m+m'+1}}\,.
\end{equation}
Then by recalling the simple statistical fact that the $n_l$s of distinct $l$ are of course independent binomial random variables,
\begin{align}
\MEAN{}{z_l^mz_{l'}^{m'}}\approx&\,\dfrac{m+m'+1}{(N_0\sum_lp_l)^{m+m'}}\MEAN{}{n_l^m\vphantom{n_{l'}^{m'}}}\MEAN{}{n_{l'}^{m'}}\nonumber\\
&\,-\dfrac{m+m'}{(N_0\sum_lp_l)^{m+m'+1}}\MEAN{}{\sum_{l''}n_{l''}n_l^mn_{l'}^{m'}}\,,
\end{align}
where
\begin{align}
\MEAN{}{\sum_{l''}n_{l''}n_l^mn_{l'}^{m'}}=&\,N_0\sum_{l''\neq l \& l'}p_{l''}\MEAN{}{n_l^m\vphantom{n_{l'}^{m'}}}\MEAN{}{n_{l'}^{m'}}\nonumber\\
&\,+\MEAN{}{n_l^{m+1}\vphantom{n_{l'}^{m'}}}\MEAN{}{n_{l'}^{m'}}+\MEAN{}{n_l^m\vphantom{n_{l'}^{m'}}}\MEAN{}{n_{l'}^{m'+1}}\,.
\end{align}

Since from App.~\ref{app:deriv}, we know that $n_l/N_0$ is a Gaussian random variable of mean $\mu_l=p_l$ and variance $\sigma_l^2=p_l(1-p_l)/N_0$ for $N_0\gg1$, the $m$th moment
\begin{equation}
\MEAN{}{n_l^m\vphantom{n_{l'}^{m'}}}=\left(N_0\dfrac{\sigma_l}{\sqrt{2}\,\I}\right)^m\HERM{m}{\I\,\dfrac{\mu_l}{\sqrt{2}\sigma_l}}
\label{eq:split}
\end{equation} 
is a simple function of the $m$th-degree Hermite polynomial $\HERM{m}{\,\cdot\,}$. Using the simple relation $\HERM{m+1}{y}=2y\,\HERM{m}{y}-2m\,\HERM{m-1}{y}$ that permits changes in the polynomial degree, we obtain the useful identity
\begin{equation}
\MEAN{}{n_l^{m+1}\vphantom{n_{l'}^{m'}}}=\mu_lN_0\MEAN{}{n_l^m\vphantom{n_{l'}^{m'}}}+(m+1)(N_0\sigma_l)^2\MEAN{}{n_l^{m-1}\vphantom{n_{l'}^{m'}}}
\end{equation}
that can now be applied to Eq.~\eqref{eq:split} to get
\begin{align}
\MEAN{}{\sum_{l''}n_{l''}n_l^mn_{l'}^{m'}}=&\,\MEAN{}{\sum_{l''}n_{l''}}\MEAN{}{n_l^m\vphantom{n_{l'}^{m'}}}\MEAN{}{n_{l'}^{m'}}\nonumber\\
&\,+(m+1)(N_0\sigma_l)^2\MEAN{}{n_l^{m-1}\vphantom{n_{l'}^{m'}}}\MEAN{}{n_{l'}^{m'}}\nonumber\\
&\,+(m'+1)(N_0\sigma_l)^2\MEAN{}{n_l^{m}\vphantom{n_{l'}^{m'}}}\MEAN{}{n_{l'}^{m'-1}}\,.
\end{align}

For sufficiently dense sampling the sum $\sum_ln_l$ should also be independent of $n_l$ since the sum is contributed by very many terms that are all distinct and therefore independent from $n_l$. Using the asymptotic relation $\HERM{m}{y}\approx(2y)^m$ for $y\gg1$, we find that
\begin{align}
\MEAN{}{z_l^mz_{l'}^{m'}}\approx&\,\MEAN{}{\dfrac{1}{(\sum_ln_{l})^{m+m'}}}\MEAN{}{n_l^m\vphantom{n_{l'}^{m'}}}\MEAN{}{n_{l'}^{m'}}\nonumber\\
&\,-\dfrac{(m+m')p_l^mp_{l'}^{m'}}{\MEAN{}{N}(\sum_lp_l)^{m+m'}}f_{m,m'}(p_l,p_{l'})\,,
\end{align}
where $f_{m,m'}(x,y)=(m+1)(1-x)+(m'+1)(1-y)$. The next order in the asymptotic expansion of $\HERM{m}{y}$ gives a smaller correction to $\MEAN{}{z_l^mz_{l'}^{m'}}$. 

We can repeat the exercise and obtain the asymptotic formulas
\begin{align}
\MEAN{}{\dfrac{n_{l'}^{m'}}{(\sum_ln_l)^m}}\approx&\,\MEAN{}{\dfrac{1}{(\sum_ln_{l})^{m}}}\MEAN{}{n_{l'}^{m'}}\nonumber\\
&\,-\dfrac{m(m'+1)(1-p_{l'})}{\MEAN{}{N}}\dfrac{(N_0p_{l'})^{m'}}{(N_0\sum_lp_l)^m}
\end{align}
and
\begin{align}
\MEAN{}{\dfrac{1}{(\sum_ln_l)^{m+m'}}}\approx&\,\MEAN{}{\dfrac{1}{(\sum_ln_l)^{m}}}\MEAN{}{\dfrac{1}{(\sum_ln_l)^{m'}}}\nonumber\\
&\,+\dfrac{2mm'}{\left(\MEAN{}{N}\right)^{m+m'+2}}\sum_lp_l(1-p_l)\,,
\end{align}
the latter is obtained from the second-order Taylor expansion of $1/(\sum_ln_l)^{m+m'}$ in $n_l$ about $N_0p_l$,
\begin{align}
\MEAN{}{\dfrac{1}{(\sum_ln_l)^{m+m'}}}\approx&\,\dfrac{1}{\left(\MEAN{}{N}\right)^{m+m'}}\nonumber\\ \nonumber\\
&\,+(N_0\sigma_l)^2\dfrac{(m+m')(m+m'+1)}{\left(\MEAN{}{N}\right)^{m+m'+2}}\,,
\end{align}
in which the first-order term vanishes since $\MEAN{}{n_l}=N_0p_l$. These relations inform us that all statistical bias are asymptotic in nature. Combining all elements and keeping terms up to first order in $1/\MEAN{}{N}$ gives
\begin{equation}
\MEAN{}{z_l^mz_{l'}^{m'}}\approx\MEAN{}{z_l^m}\MEAN{}{z_{l'}^{m'}}+O\!\left(\dfrac{1}{\MEAN{}{N}(\sum_lp_l)^{m+m'}}\right)\,.
\end{equation}
%\begin{equation}
%\MEAN{}{z_l^mz_{l'}^{m'}}\approx\MEAN{}{z_l^m}\MEAN{}{z_{l'}^{m'}}+\dfrac{p^m_lp^{m'}_{l'}}{\MEAN{}{N}(\sum_lp_l)^{m+m'}}\left[2mm'\dfrac{\sum_lp_l(1-p_l)}{\sum_lp_l}-m(m'+1)(1-p_{l'})-m'(m+1)(1-p_{l})\right]\,.
%\end{equation}
Finally, invoking Kac's theorem confirms asymptotic independence between $z_l$ and $z_{l'}$, and thereafter for the whole set $\{z_l\}$ of these UHOM random variables.

\section{Realistic detections}
\label{app:real}

It is a simple matter to show that the first main result remains unchanged for realistic detections. Suppose that all photodetectors now have the efficiency $0\leq\eta\leq1$. Then standard characteristic-function treatment~(see for instance \cite{Lai:1989ah}) allows us to find that the more realistic measured outcomes for HET are, instead of the usual coherent states $\ket{\alpha}\bra{\alpha^*}$, given by the full-rank statistical mixtures
\begin{align}
&\,\dfrac{\eta}{1-\eta}\int\dfrac{(\D\alpha')}{\pi}\ket{\alpha'}\E{-\frac{\eta}{1-\eta}|\alpha-\alpha'|^2}\bra{\alpha'^*}\nonumber\\
=&\,\eta\bm{:}\E{-\eta(a^\dagger-\alpha^*)(a-\alpha)}\bm{:}\,.
\end{align}
Alternatively, Born's rule dictates that the realistic HET setup is equivalently the perfect HET setup with the quantum state $\rho$ transformed to $\rho'$ by a corresponding Gaussian twirling operation. Then, the expression $\sCRB_\textsc{het}'$ in (3) of the main text can be obtained by the simple replacement $\rho\rightarrow\rho'$.

For UHOM, the results in \cite{Wallentowitz:1996qo} show that the binomial probability for ``no-click'' detections is transformed to $p(\alpha,\eta)=\left<\bm{:}\E{-\eta(a^\dagger-\alpha^*)(a-\alpha)}\bm{:}\right>$. Furthermore, in going from the discretized sum to the continuous integral limit (review Sec.~\ref{app:deriv} of this SM), we note that $\sum_lp'_l=\sum_lp(\alpha_l,\eta)\rightarrow\pi/[(\D\alpha)\eta]$, which contributes the multiplicative factor $\eta$ in the expression for $\sCRB_\textsc{uhom}'$.

\section{Moment tomography}
\label{app:momtom}

It is easily verified that for the $m$th operator moment ($m\geq l$) that is Weyl ordered in the position $X$ and momentum $P$, its corresponding Wigner function is given by $x^lp^{m-l}$ in terms of the phase-space variables $x$ and $p$. There is then a simple one-to-one relation between $\rvec{v}_\textsc{p}(\alpha)$ and $\rvec{v}_\textsc{w}(\alpha)$ as a consequence of the Gauss transform. These are given by
\begin{align}
\rvec{v}_\textsc{w}(\alpha)\,\widehat{=}\begin{pmatrix}
x\\
p
\end{pmatrix}&\leftrightarrow\rvec{v}_\textsc{p}(\alpha)\,\widehat{=}\begin{pmatrix}
x\\
p
\end{pmatrix}\nonumber\\
\rvec{v}_\textsc{w}(\alpha)\,\widehat{=}\begin{pmatrix}
x^2\\
x p\\
p^2
\end{pmatrix}&\leftrightarrow\rvec{v}_\textsc{p}(\alpha)\,\widehat{=}\begin{pmatrix}
x^2-\frac{1}{2}\\
x p\\
p^2-\frac{1}{2}
\end{pmatrix}\nonumber\\
\rvec{v}_\textsc{w}(\alpha)\,\widehat{=}\begin{pmatrix}
x^3\\
x^2p\\
x p^2
p^3
\end{pmatrix}&\leftrightarrow\rvec{v}_\textsc{p}(\alpha)\,\widehat{=}\begin{pmatrix}
x^3-\frac{3}{2}x\\
x^2p-\frac{1}{2}p\\
x p^2-\frac{1}{2}x\\
p^3-\frac{3}{2}p
\end{pmatrix}\nonumber\\
\rvec{v}_\textsc{w}(\alpha)\,\widehat{=}\begin{pmatrix}
x^4\\
x^3 p\\
x^2p^2\\
x p^3\\
p^4
\end{pmatrix}&\leftrightarrow\rvec{v}_\textsc{p}(\alpha)\,\widehat{=}\begin{pmatrix}
x^4-3x^2+\frac{3}{4}\\
x^3p-\frac{3}{2}xp\\
x^2p^2-\frac{1}{2}x^2-\frac{1}{2}p^2+\frac{1}{4}\\
x p^3-\frac{3}{2}xp\\
p^4-3p^2+\frac{3}{4}
\end{pmatrix}\,.
\label{eq:ident}
\end{align}
Then the evaluation of $\sCRB_\textsc{bhomopt}$, $\sCRB_\textsc{het}$ and $\sCRB_\textsc{uhom}$ amounts to the evaluation of all integrals involving $\rvec{v}_\textsc{p}(\alpha)$ and $\rvec{v}_\textsc{w}(\alpha)$ using the identities in \eqref{eq:ident}. 

For $\sCRB_\textsc{het}$ and $\sCRB_\textsc{uhom}$, this can be easily accomplished with the help of characteristic functions $\displaystyle\chi_1=\overline{\E{g^*\alpha+g\alpha^*}}\,\,\left[g=(u+\I v)/\sqrt{2}\right]$ and $\displaystyle\chi_2=\overline{\overline{\E{g^*\alpha+g\alpha^*}}}$, where the single and double overlines respectively denote the phase-space integrals with respect to $\QA{\alpha}$ and (unnormalized) $\QA{\alpha}\left[1-\QA{\alpha}\right]$. Then 
\begin{align}
\overline{x^kp^l}&=\left(\dfrac{\partial}{\partial u}\right)^k\left(\dfrac{\partial}{\partial v}\right)^l\chi_1\Bigg|_{u,v=0}\,,\nonumber\\
\overline{\overline{x^kp^l}}&=\left(\dfrac{\partial}{\partial u}\right)^k\left(\dfrac{\partial}{\partial v}\right)^l\chi_2\Bigg|_{u,v=0}
\end{align}
supply the required quantities.

The centralized Gaussian states of the Husimi-function covariance matrix $\dyadic{G}_\textsc{het}$ have the characteristic functions
\begin{align}
\chi_1&=\exp\!\left(\frac{\DET{\dyadic{G}_\textsc{het}}}{2}\rvec{g}^\dagger\dyadic{M}\,\rvec{g}\right)\,,\nonumber\\
\chi_2&=\frac{1}{2\sqrt{\DET{\dyadic{G}_\textsc{het}}}}\exp\!\left(\frac{\DET{\dyadic{G}_\textsc{het}}}{4}\rvec{g}^\dagger\dyadic{M}\,\rvec{g}\right)\,,
\end{align} 
where $\rvec{g}\,\widehat{=}\,\TP{(-g\,\,\,g^*)}$, $\dyadic{M}=\dyadic{H}^\dagger\dyadic{G}_\textsc{het}^{-1}\,\dyadic{H}$, and $\dyadic{H}\,\widehat{=}\dfrac{1}{\sqrt{2}}\begin{pmatrix}
1 & 1\\
-\I & \I
\end{pmatrix}$. Those of the Fock states read
\begin{align}
\chi_1&=\E{|g|^2}\,\LAG{n}{-|g|^2}\,,\nonumber\\
\chi_2&=\frac{1}{2^{2n+1}}\binom{2n}{n}\E{\frac{|g|^2}{2}}\,\LAG{2n}{-\frac{|g|^2}{2}}\,.
\end{align}

For the setting $\mu=\lambda$, the $\sCRB$ expressions are catalogued as follows:
\begin{align}
\sCRB_\textsc{1,het}=&\,\frac{1}{2}(3+\mu^2)\,,\nonumber\\
\sCRB_\textsc{1,uhom}=&\,\sCRB_\textsc{1,het}-\frac{3+\mu ^2}{4\sqrt{2+2\mu ^2}}\,,\nonumber\\
\sCRB_\textsc{2,het}=&\,\frac{1}{2}(6+3\mu^2+\mu^4)\,,\nonumber\\
\sCRB_\textsc{2,uhom}=&\,\sCRB_\textsc{2,het}-\dfrac{17+8 \mu ^2+3 \mu ^4}{16 \sqrt{2+2\mu ^2}}\,,\nonumber\\
\sCRB_\textsc{3,het}=&\,\frac{1}{8}\left(85+35\mu^2+33\mu^4+15\mu^6\right)\,,\nonumber\\
\sCRB_\textsc{3,uhom}=&\,\sCRB_\textsc{3,het}-\frac{77+21\mu^2 +15\mu^2+15\mu^4}{64\sqrt{2+2\mu^2}}\,,\nonumber\\
\sCRB_\textsc{4,het}=&\,\frac{1}{8}\left(396+117 \mu ^2+148 \mu ^4+135 \mu ^6+48 \mu ^8\right)\,,\nonumber\\
\sCRB_\textsc{4,uhom}=&\,\sCRB_\textsc{4,het}\nonumber\\
&\,-\dfrac{735+142 \mu ^2+40 \mu ^4+234 \mu ^6+177 \mu ^8}{256\sqrt{2+2\mu^2}}\,.
\end{align}
The corresponding expressions for the Fock states are given by
\begin{align}
\sCRB_\textsc{1,het}=&\,2(n+1)\,,\nonumber\\
\sCRB_\textsc{1,uhom}=&\,\sCRB_\textsc{1,het}-\frac{\Gamma \left(n+\frac{3}{2}\right)}{\sqrt{\pi}\,\Gamma (n+1)}\,,\nonumber\\
\sCRB_\textsc{2,het}=&\,\frac{1}{2}(n+1)(3n+10)\,,\nonumber\\
\sCRB_\textsc{2,uhom}=&\,\sCRB_\textsc{2,het}-\binom{2 n}{n}\dfrac{(n+1) (6 n+7)}{2^{2 n+3}} \,,\nonumber\\
\sCRB_\textsc{3,het}=&\,(n+1) \left(6 n^2+20 n+21\right)\,,\nonumber\\
\sCRB_\textsc{3,uhom}=&\,\sCRB_\textsc{3,het}-\frac{(6n^2+5n+4)\,\Gamma \left(n+\frac{3}{2}\right)}{2 \sqrt{\pi }\,\Gamma (n+1)}\,,\nonumber\\
\sCRB_\textsc{4,het}=&\,\frac{1}{8} (n+1) (45 n^3+437 n^2+1040 n+844)\,,\nonumber\\
\sCRB_\textsc{4,uhom}=&\,\sCRB_\textsc{4,het}-(n+1)\nonumber\\
&\,\times\frac{(180n^3+544n^2+521n+166)\, \Gamma \left(n+\frac{1}{2}\right)}{64 \sqrt{\pi }\, \Gamma (n+1)}\,.
\end{align}

The $\sCRB_\textsc{bhomopt}$ expressions for the improved strategy of HOM can be found in an analogous way by looking at the operator moments and calculating the Fisher information matrix~\cite{Teo:2017aa}. Further analysis shall be reported elsewhere but for now, we simply supply all the final analytical results that are obtainable from the theory. These are
\begin{align}
\sCRB_\textsc{1,bhomopt}=&\,\frac{1}{2} (1+\mu)^2\,,\nonumber\\
\sCRB_\textsc{2,bhomopt}=&\,\frac{1}{4} \left(2+5 \mu+2 \mu ^2+5 \mu ^3 +2 \mu ^4\right)\,,\nonumber\\
\sCRB_\textsc{3,bhomopt}=&\,\frac{5}{24} (9+30 \mu+9 \mu ^2+16 \mu ^3+9 \mu ^4\nonumber\\
&\,\quad\,\,\,+30 \mu ^5+9 \mu ^6)\,,\nonumber\\
\sCRB_\textsc{4,bhomopt}=&\,6+\frac{1}{6} \mu \left(\mu ^2+1\right) \nonumber\\
&\,\times\left(153 + 36 \mu - 88 \mu^2 + 153 \mu^4 + 36 \mu^5\right)\nonumber\\
\end{align}
for the Gaussian states, and
\begin{align}
\sCRB_\textsc{1,bhomopt}=&\,2 (2 n+1)\,,\nonumber\\
\sCRB_\textsc{2,bhomopt}=&\,4 \left(n^2+n+1\right)\,,\nonumber\\
\sCRB_\textsc{3,bhomopt}=&\,\frac{14}{9} \left(20 n^3+30 n^2+40 n+15\right)\,,\nonumber\\
\sCRB_\textsc{4,bhomopt}=&\frac{77}{36}\left(17 n^4 + 34 n^3+ 139 n^2 + 122 n+48  \right)\nonumber\\
\end{align}
for the Fock states.

\end{document}